\documentclass[aps,prb,10pt,twocolumn,showpacs,groupedaddress,superscriptaddress]{revtex4-2}
\usepackage{amssymb}
\usepackage{eurosym}
\usepackage{dcolumn}
\usepackage{bm}
\usepackage{graphicx}
\usepackage{color}
\usepackage{amsmath}
\usepackage{amsfonts}
\usepackage{lmodern}
\usepackage[bookmarks=false]{hyperref}

\begin{document}
\title{Low-energy spin excitations of the frustrated ferromagnetic $J_1$-$J_2$ chain material linarite, PbCuSO$_4$(OH)$_2$, in applied magnetic fields ${\bf H} \parallel b$ axis}
\author{L. Heinze}
\email{Corresponding author: l.heinze@tu-braunschweig.de}
\affiliation{Institut f\"ur Physik der Kondensierten Materie, Technische Universit\"at Braunschweig, 38106 Braunschweig, Germany}
\author{M. D. Le}
\affiliation{ISIS Neutron and Muon Source, Rutherford Appleton Laboratory, Chilton, Didcot, OX11 0QX, United Kingdom}
\author{O. Janson}
\affiliation{Leibniz Institute for Solid State and Materials Research IFW Dresden, 01069 Dresden, Germany}
\author{S. Nishimoto}
\affiliation{Leibniz Institute for Solid State and Materials Research IFW Dresden, 01069 Dresden, Germany}
\affiliation{Department of Physics, Technical University Dresden, 01062 Dresden, Germany}
\author{A. U. B. Wolter}
\affiliation{Leibniz Institute for Solid State and Materials Research IFW Dresden, 01069 Dresden, Germany}
\author{S. S\"{u}llow}
\affiliation{Institut f\"ur Physik der Kondensierten Materie, Technische Universit\"at Braunschweig, 38106 Braunschweig, Germany}
\author{K. C. Rule}
\affiliation{Australian Nuclear Science and Technology Organisation, Lucas Heights, NSW 2234, Australia}
\date{\today}

\begin{abstract}
We report a study of the spin dynamics of the frustrated ferromagnetic $J_1$-$J_2$ chain compound linarite, PbCuSO$_4$(OH)$_2$, in applied magnetic fields up to field polarization. By means of an extreme-environment inelastic neutron scattering experiment, we have measured the low-energy spin excitations of linarite in fields up to 8.8\,T for $\mathbf{H} \parallel b$ axis. We have recorded the spin excitation spectra along $h$, $k$ and $l$ for the field-induced magnetic phases IV, V and the field polarized state close to saturation. By employing first-principles calculations, we estimate the leading magnetic exchanges out of the $bc$ plane and model the dispersion relations using linear spin-wave theory. In this way, we find evidence for a (very weak) residual magnetic exchange coupling out of the $bc$ plane. Together with the previously established dominant intrachain couplings $J_1$ and $J_2$ and the interchain coupling $J_3$, we derive an effective set of exchange couplings for a microscopic description of linarite. Further, we find that the peculiar character of phase V manifests itself in the measured spin dynamics.
\end{abstract}

\maketitle

\section{Introduction}
\label{sec:intro}

Low-dimensional quantum magnetic systems are fascinating subjects of study both from an experimental as well as from a theoretical point of view. For experimental studies, in particular the material class of copper oxides plays an important role. Here, the copper ions carry a spin $S = 1/2$ and the magnetic exchange dominantly arises from superexchange via the oxygen orbitals. The exchange interactions can be of a competing nature and magnetic frustration may arise~\cite{Lacroix2011} leading to a multitude of exotic ground states with highly interesting spin dynamics (see for instance Refs.~\cite{Broholm2020,Rule2011,Yoshida2017,Zhang2020}). 

More specifically, in (quasi) one-dimensional systems, magnetic frustration can occur due to competing exchange interactions along a spin chain when there is also a relevant next-nearest neighbor exchange $J_2$ in addition to the nearest neighbor exchange $J_1$. The corresponding Hamiltonian reads

\begin{equation}
\mathcal{H} = J_1 \sum_i \mathbf{S}_i\,\mathbf{S}_{i+1} + J_2 \sum_i \mathbf{S}_i\,\mathbf{S}_{i+2} - h \sum_i S_i^z
\label{eq:Hamiltonian}
\end{equation}

\noindent for a bare chain in a magnetic field $h$ with $\mathbf{S}_i$ being the spin-1/2 operator at chain site $i$. This model has been studied in great detail and especially the {\it frustrated ferromagnetic} ({\it ff}) $J_1$-$J_2$ chain (ferromagnetic $J_1 < 0$ and antiferromagnetic $J_2 > 0$) has attracted considerable theoretical interest (see for instance Refs.~\cite{Chubukov1991,Hikihara2008,Sudan2009,HeidrichMeisner2009,Zhitomirsky2010,Onishi2015,Balents2016}). Its magnetic phase diagram~\cite{Hikihara2008,Sudan2009,HeidrichMeisner2009}, depending on the ratio $\alpha = J_1/J_2$, contains exotic field-induced states such as spin density wave (SDW) and spin-multipolar phases. The latter arises from a multi-magnon condensation close to saturation. For zero/low magnetic fields a vector chiral phase is predicted. The model in Eq.~(\ref{eq:Hamiltonian}) can be extended by including, for instance, interchain or Dzyaloshinskii-Moriya (DM) interaction~\cite{Jinterchain_and_DM}.

While some of the above mentioned states have been identified in real materials, the challenging field and temperature ranges required can make an experimental test cumbersome. Moreover, especially spin-multipolar states are notoriously hard to be accessed experimentally. From a theoretical viewpoint, ``experimental signatures'' of a quadrupolar state (also referred to as spin-nematic) have been identified for measurement techniques such as nuclear magnetic resonance (NMR)~\cite{Sato2009} and electron spin resonance~\cite{Furuya2017}. The low-energy spin excitation spectra of a {\it ff} $J_1$-$J_2$ chain~\cite{Onishi2015,Flavia2018} can be examined to some extent using inelastic neutron scattering (INS).

Over the years, a few materials have been established as model systems of the {\it ff} $J_1$-$J_2$ chain. Particular interest lies on experimental realizations with $-4 < \alpha < 0$. This is the range of $\alpha$ where, \textit{e.g.}, the compounds LiCuVO$_4$~\cite{Enderle2005,Enderle2010,Nishimoto2011}, PbCuSO$_4$(OH)$_2$ (linarite)~\cite{Wolter2012,Rule2017,Cemal2018} and $\beta$-TeVO$_4$~\cite{Saul2014,Weickert2016} were established to reside, and which have been studied actively during the past decade. The spin dynamics in applied fields have been investigated for a few of them by means of NMR and INS~\cite{Buettgen2014,Orlova2017,Heinze2019,Pregelj2020}, in particular in view of spin-multipolar states. For these investigations the application of high magnetic fields (a few 10 Tesla) is essential---an obstacle for the study of such materials. Still, in Ref.~\cite{Buettgen2014}, a high-field NMR study on LiCuVO$_4$ was reported, from which a narrow field range of $\sim 1$\,T close to saturation ($\sim 41$\,T) was identified to possibly host a spin-nematic state. Additional high-field NMR measurements to test this scenario followed~\cite{Orlova2017}.

The object of the present study is the mineral linarite, which has been extensively characterized regarding its magnetic properties within the past ten years~\cite{Wolter2012,Rule2017,Cemal2018,Willenberg2012,Schaepers2013,Schaepers2014,Willenberg2016,Povarov2016,Feng2018,Heinze2019,Gotovko2019,Gillig2021}. Its anisotropic magnetic phase diagram is well established. With a N\'eel temperature $T_\mathrm{N} = 2.8$\,K and a saturation field of $\mu_0H_\mathrm{sat} = 9.64$\,T ($\mathbf{H} \parallel b$ axis)~\cite{Wolter2012,Heinze2019} it is comparatively easy to access. This allows to use static fields $\sim \mu_0H_\mathrm{sat}$ and to apply INS to record the spin excitation spectra of a {\it ff} $J_1$-$J_2$ chain just below saturation. 

A theoretical proposal for an experimental observation of a spin-nematic state in a frustrated ferromagnet by means of INS is given in Ref.~\cite{Smerald2015} for the (quasi)-2D magnet BaCdVO(PO$_4$)$_2$. Here, it is predicted that a (ghostly) linearly dispersing Goldstone mode might be observable close to saturation. Applied to a {\it ff} $J_1$-$J_2$ chain chain with interchain couplings and in the presence of a 2D spin-nematic state, similarly, a linearly dispersing Goldstone mode should be observable~\cite{Starykh2014}. However, given that its spectral weight might be minute compared to the dispersion of the one-magnon excitations~\cite{Smerald2015}, it is still extremely challenging to resolve it. In spite of this, here, we report a corresponding in-field INS study on the field induced magnetic phases of linarite up to the field polarized, paramagnetic state. From a previous NMR study on linarite, two-magnon excitations were found to be the lowest-lying spin excitations in the high-field region of the phase diagram~\cite{Heinze2019}. This indicates the possibility of spin-nematicity ($p=2$) whereas the $\alpha$ value of linarite would rather be indicative of a multipolar state formed by $p>2$ bound magnons. Hence, if the Goldstone mode can be revealed remains to be seen.

\begin{figure}[t!]
\includegraphics[width=\linewidth]{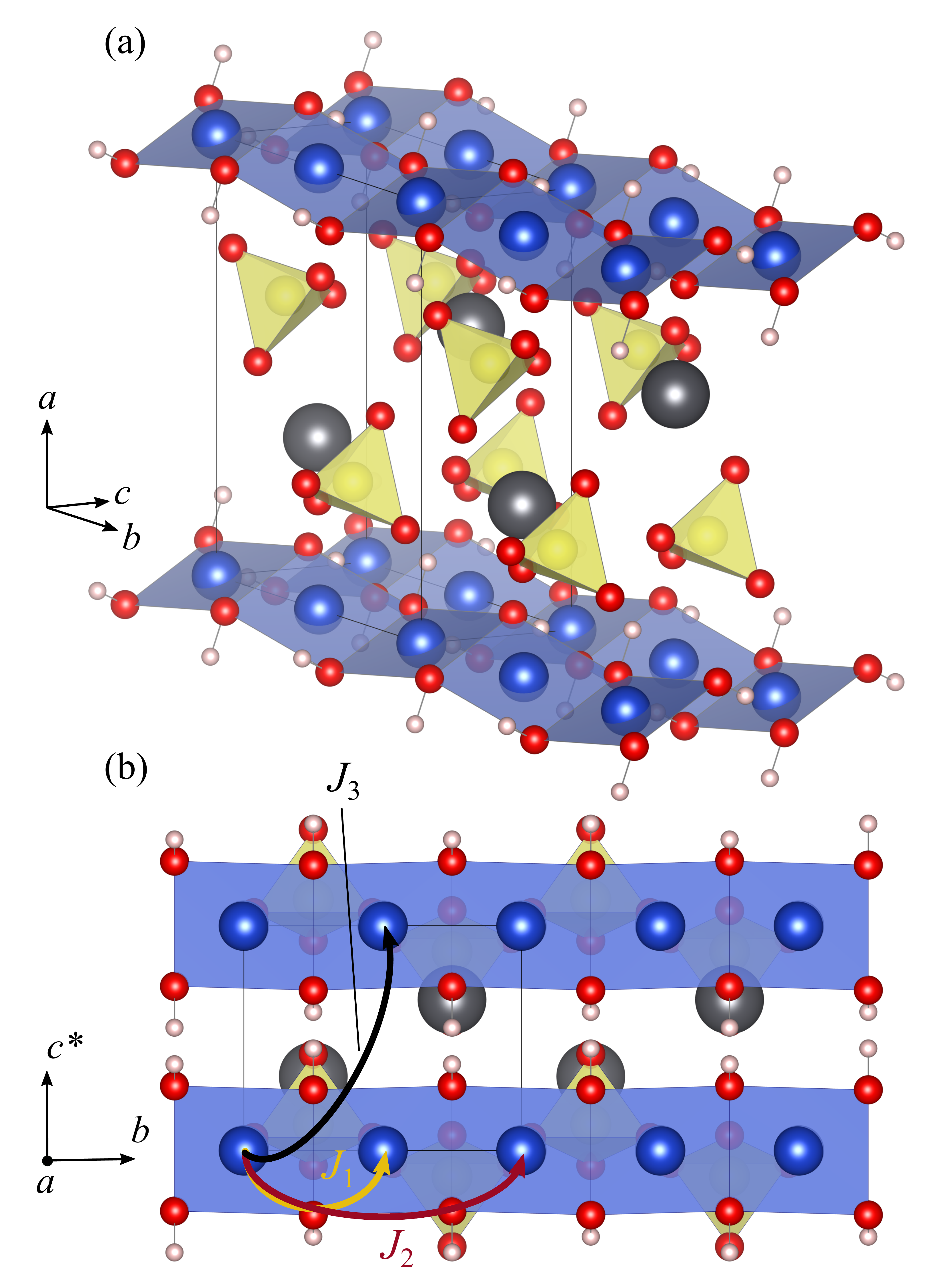}
\includegraphics[width=\linewidth]{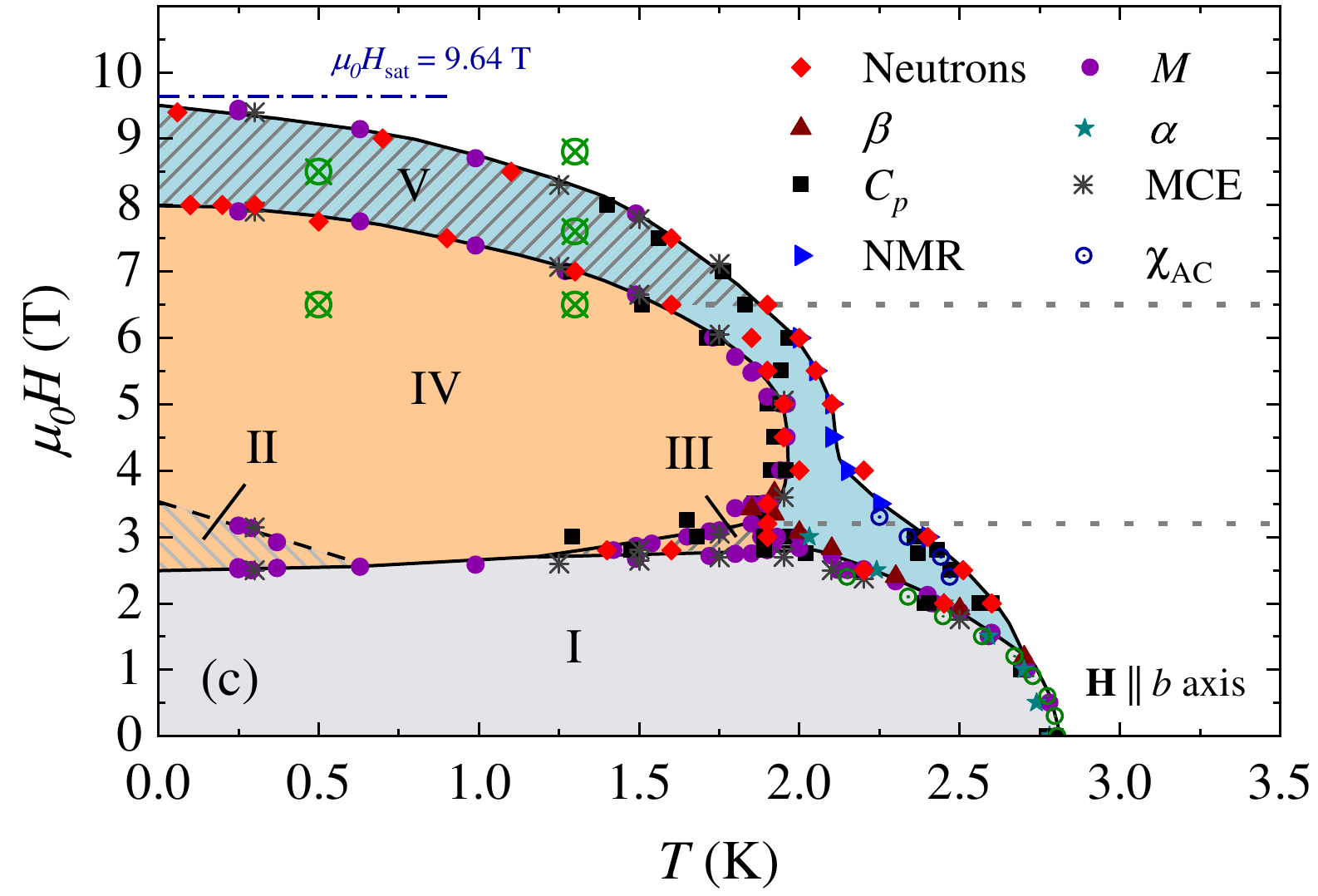}
\caption{(a)--(b) Crystal structure of linarite, PbCuSO$_4$(OH)$_2$, with Cu (blue), Pb (black), S (yellow), O (red), H (white)~\cite{Schaepers2013}. In (b), the view on the $bc^*$ plane is displayed for a visualization of the magnetic coupling scheme consisting of $J_1$, $J_2$ and $J_3$, as modeled in Ref.~\cite{Rule2017}. The solid lines indicate the structural unit cell. (c) Magnetic phase diagram of linarite for $\mathbf{H} \parallel b$ axis consisting of phases/regions I--V; green circles with crosses indicate where the INS measurements were carried out.}
\label{fig:phasediagram}
\end{figure}

Linarite, PbCuSO$_4$(OH)$_2$, crystallizes in a monoclinic structure with space group $P2_1/m$~\cite{Schofield2009,Effenberger1987,Schaepers2013} with $a = 9.682$, $b = 5.646$, $c = 4.683$\,\AA~and $\beta = 102.66^{\circ}$~\cite{Schofield2009} (Fig.~\ref{fig:phasediagram}~(a)--(b)). Herein, buckled, edge-sharing Cu(OH)$_4$ units form chains along the $b$ axis. Initially, the magnetic exchange in linarite has been modeled with two dominant exchange couplings $J_1 = -100$\,K and $J_2 = 36$\,K ($\alpha = -2.78$) via bulk measurements~\cite{Wolter2012}. Subsequently, these exchange couplings were confirmed in a zero-field INS study~\cite{Rule2017}: In combination with an analysis by means of linear spin-wave theory (LSWT), this study yielded $J_1 = -114$\,K and $J_2 = 37$\,K ($\alpha = -3.08$) as well as an additional (effective) interchain coupling $J_\mathrm{ic} \equiv J_3 = 4$\,K (Fig.~\ref{fig:phasediagram}~(b)). These findings were well supported by dynamical density-matrix renormalization group (DDMRG) results with $J_1 = -78$\,K, $J_2 = 28$\,K ($\alpha = -2.79$) and $J_3 = 7$\,K~\cite{Rule2017}. From a subsequent in-field INS study~\cite{Cemal2018}, it was concluded on the existence of three additional couplings aside from the previously established $J_1$--$J_3$. Despite the limited set of INS data, it was inferred that linarite was closer to the quantum critical point $\alpha=-4$ with $J_1 = -168$\,K, $J_2 = 46$\,K ($\alpha = -3.65$) and $J_3 = 8$\,K.

The competing interactions in linarite are responsible for its complex magnetic phase diagram~\cite{Wolter2012,Willenberg2012,Schaepers2013,Willenberg2016,Povarov2016,Cemal2018,Feng2018,Heinze2019,Gotovko2019}. A helical magnetic ground state was observed below $T_\mathrm{N}$ with an incommensurate magnetic propagation vector of ${\bf q}$ = (0, 0.186, 0.5)~\cite{Willenberg2012}. The field-induced behavior of linarite is highly sensitive to the field direction and the temperature~\cite{Schaepers2013,Feng2018,Cemal2018,Gotovko2019}. By now, the phase diagram of linarite has been studied by a multitude of experimental techniques for $\mathbf{H} \parallel b$ axis (Fig.~\ref{fig:phasediagram}~(c)). In this field configuration, the helical ground-state phase (labeled I) is suppressed in low magnetic fields of 2.5--3\,T. It is replaced by an antiferromagnetic phase IV with a commensurate propagation vector $\mathbf{q}$ = (0, 0, 0.5)~\cite{Willenberg2016}. At lowest temperatures, $T \lesssim 600$\,mK, a hysteretic region II separates phases I and IV~\cite{Schaepers2013,Heinze2019}. Above $\sim 1.2$\,K, a coexistence phase III is present between phases I and IV~\cite{Willenberg2016}. The phases I, III and IV are enveloped by a phase V which has been refined as a incommensurate, longitudinal SDW phase with $\mathbf{q}$ = (0, $k_y$, 0.5)~\cite{Willenberg2016}. Here, $k_y$ shifts as function of field and temperature~\cite{Willenberg2016,Heinze2019}.

In the present work, we advance the study of the spin dynamics of linarite. Motivated by the search for spin-multipolar states, we have carried out time-of-flight neutron scattering measurements in magnetic fields up to 8.8\,T for ${\bf H} \parallel b$ axis and temperatures as low as 500\,mK. From our data, we extract the low-energy spin excitation spectra within the magnetic phases IV, V and for the field polarized state close to saturation. Given the very good agreement between experiment and theory for the INS study in zero magnetic field, we first model the INS data using DDMRG and LSWT starting with the $J$ parameters from Ref.~\cite{Rule2017}. However, this model lacks exchange couplings out of the $bc$ plane that are a prerequisite for modeling the experimentally measured dispersions along $h$. To remedy this shortcoming, we perform first-principles calculations of linarite and evaluate the respective transfer integrals by mapping the relevant part of the band structure onto a Wannier basis of Cu $3d_{x^2-y^2}$ states. We find evidence for a weak coupling out of the $bc$ plane, $J_4$, which is further corroborated by the LSWT results. In view of the applied-field spin-wave analysis, we discuss our INS data regarding the possibility of a spin-multipolar state close to saturation.

\section{Experimental details}
\label{sec:experiment}

Inelastic neutron scattering experiments on linarite were carried out at the ISIS Neutron and Muon Source of the Rutherford Appleton Laboratory using the cold neutron multi-chopper spectrometer LET~\cite{LET}. Due to the small sample size of the naturally grown single crystals, a crystal array with a total mass of $\sim 150$\,mg was prepared for the experiment. The crystals were sourced from the Blue Bell Mine, San Bernardino, USA, and had also been used for the previous INS study on linarite~\cite{Rule2017}. Each crystal was needle-shaped and separately characterized using neutron diffraction whereupon the long axis of each crystal was identified as $b$ axis. A set of five high-quality single crystals without twins and with low mosaic spread ($\mathrm{FWHM} < 2^\circ$) was co-aligned on thin aluminum plates. The crystals were affixed using hydrogen-free Teflon tape to ensure low background scattering~\cite{Rule2018glue}. The $b$ axis was aligned vertical such that the $a$ and the $c$ axes were contained within the scattering plane. This enabled recording of the spin excitations along $h$ and $l$ (perpendicular to the chain direction). The large position sensitive detectors of LET further allowed coverage of a $k$ range up to $\pm 0.3$\,r.l.u.~($E_i=2.981$\,meV) out of the scattering plane to record the spin excitations along the chain direction.

For the experiment at the LET instrument~\cite{RB1910225}, a $^3$He insert was used in combination with the wide-bore, 9\,T vertical-field magnet in order to carry out the neutron scattering scans in the required temperature region of the magnetic phase diagram. The magnetic field pointed along the crystallographic $b$ axis. By keeping $T=1.3$\,K fixed, the magnetic phases I, IV, V and the field polarized state could be reached by varying field only (Fig.~\ref{fig:phasediagram}~(c)). LET was operated in high-flux mode with chopper frequencies of $f_1 = 40$\,Hz, $f_2 = 10$\,Hz, $f_3 = 80$\,Hz (pulse removal), $f_4 = 120$\,Hz, $f_5 = 240$\,Hz (resolution) in order to provide in parallel incident neutron energies $E_i =$ 8.246, 2.981, 1.518, and 0.918\,meV. The focus of the present analysis lies on data set with $E_i = 2.981$\,meV affording an energy resolution of 0.11\,meV.

The sample was zero-field cooled to 1.3\,K, then a magnetic field was applied. Neutron scattering scans were carried out at 1.3\,K in fields of 8.8\,T (beyond phase V), 7.6\,T (phase V) and 6.5\,T (phase IV), measuring from highest to lowest field while keeping the temperature fixed. A corresponding data set was recorded at 500\,mK in magnetic fields of 8.5\,T (phase V) and 6.5\,T (phase IV)~\cite{SI}. In the present study, the data analysis is focused primarily on the data set collected at 1.3\,K, which enables investigation of the magnetic state beyond phase V. In Fig.~\ref{fig:phasediagram}~(c), the ($T$, $\mu_0 H$) points of the neutron scattering scans are indicated by green circles with crosses. An additional scan was performed at 40\,K and 0\,T, \textit{i.e.}, for $T \gg T_\mathrm{N}$. An angular range of 90$^{\circ}$ with a step width of 1$^{\circ}$ was covered for each scan. Each step was counted for an integrated proton current of 10\,$\mu$A$\cdot$h, corresponding on average to about 15 minutes. The INS data were reduced using the MantidPlot software package~\cite{Mantid}. Slices through the measured data $I(\mathbf{Q},E)$ were carried out using the program package Horace in Matlab~\cite{Horace}.

\section{Experimental results}
\label{sec:ins_results}

\begin{turnpage}
\begin{figure*}[p]
\includegraphics[width=\textheight]{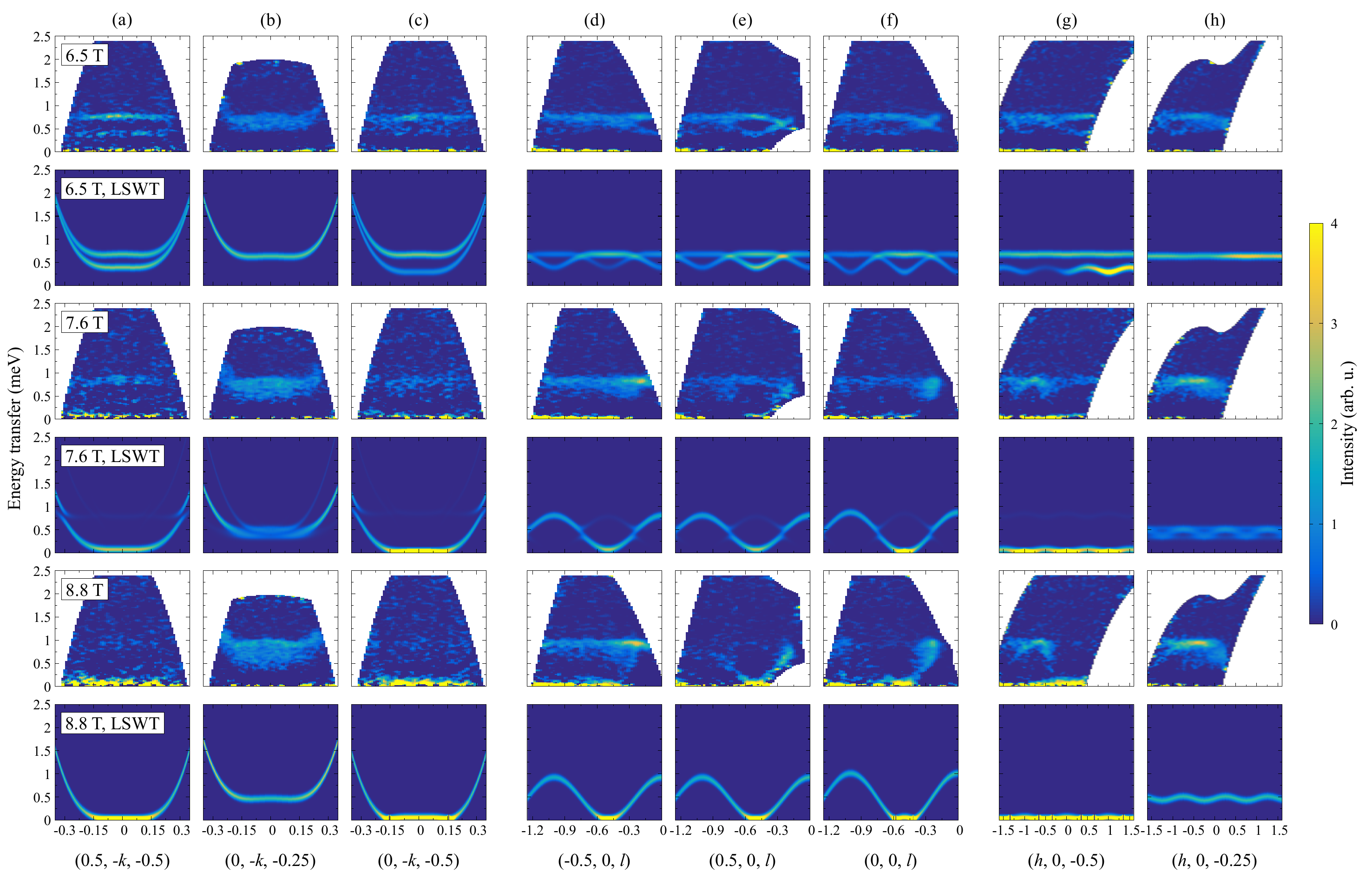}
\caption{Inelastic neutron scattering intensity maps, plotted as energy transfer over $(h, k, l)$. The experimental data were recorded at 1.3\,K in applied magnetic fields of 6.5\,T (first row), 7.6\,T (third row) and 8.8\,T (fifth row) with an incident energy $E_i=2.981$\,meV. In the second, fourth and sixth rows, the respective calculated spectra on the basis of LSWT are shown (for details see Sec.~\ref{sec:LSWT}). Reciprocal directions are given in r.l.u. For details see text.}
\label{fig:INSdata_LSWT}
\end{figure*}
\end{turnpage}

A summary of our INS data recorded at 1.3\,K with an incident energy $E_i=2.981$\,meV is displayed in Fig.~\ref{fig:INSdata_LSWT}, where we plot color-coded intensity maps as energy transfer over $\mathbf{Q}=(h, k, l)$ for three applied magnetic fields: 6.5\,T (first row), 7.6\,T (third row) and 8.8\,T (fifth row). In each column (a)--(h), the same cut through the data has been performed for the different field points, respectively. The data were smoothed over a width of two bins along the horizontal and the vertical directions. All data have been corrected for a non-magnetic background by subtracting the data measured at 40\,K in 0\,T.

In Fig.~\ref{fig:INSdata_LSWT}~(a)--(c), we present $E$-$\mathbf{Q}$ slices along $k$, {\it i.e.}, along the (reciprocal) chain direction. With a limited $k$ range of $\pm 0.3$\,r.l.u.~out of the scattering plane, we only observe a part of the spin-wave dispersion of linarite. Along ($0.5$, $-k$, $-0.5$) (Fig.~\ref{fig:INSdata_LSWT}~(a)), at 6.5\,T, we observe two gapped modes, which appear almost flat within the limited $k$ range. At 7.6\,T, scattering intensity can be observed at $\sim 0.8$\,meV energy transfer. Generally, the scattering intensity here is reduced. At 8.8\,T, we observe increased scattering intensity close to the elastic line only suggesting the presence of one spin-wave mode. This is in accordance with the sample being (almost) field-polarized in 8.8\,T. For the cuts along ($0$, $-k$, $-0.25$) (Fig.~\ref{fig:INSdata_LSWT}~(b)), at 6.5\,T, we observe dispersion with shallow minima at $k \sim \pm 0.15$. Due to the fact that we clearly observed two modes along ($0.5$, $-k$, $-0.5$) at 6.5\,T, the $E$-$\mathbf{Q}$ slice along ($0$, $-k$, $-0.25$) suggests that the two modes here lie very close together or on top of each other. At 7.6\,T, we find a broadening and intensifying of the dispersion---here likely consisting of several modes. At 8.8\,T, close to the saturation, we find the single spin-wave mode intensified and strongly broadened~\cite{cut_0k-0p25}. For the cuts along ($0$, $-k$, $-0.5$) (Fig.~\ref{fig:INSdata_LSWT}~(c)), at 6.5\,T, the two modes apparently lie close to each other. They appear dispersionless within the experimental $k$ range. Similar as for the cut along ($0.5$, $-k$, $-0.5$), at 7.6\,T, we observe weak scattering intensity at an energy transfer of $\sim 0.8$\,meV along ($0$, $-k$, $-0.5$). At 8.8\,T, we observe scattering intensity merely close to the elastic line. 

Next, we analyze the spin excitation spectra along $l$ (Fig.~\ref{fig:INSdata_LSWT}~(d)--(f)), which allow the investigation of the interchain couplings perpendicular to the chain. In accordance with Ref.~\cite{Rule2017}, we observe dispersive spin-wave modes. Along ($-0.5$, $0$, $l$) (Fig.~\ref{fig:INSdata_LSWT}~(d)), at 6.5\,T, we observe two weak dispersing modes, which are likely to cross at $l=-0.25$. These two modes are gapped. At 7.6\,T, the dominant feature is a flat band at $\sim 0.8$\,meV energy transfer with increasing spectral weight when $l \to 0$. This dispersionless feature is still present at 8.8\,T but with additional scattering intensity close to the elastic line at $l=-0.5$. Along ($0.5$, $0$, $l$) (Fig.~\ref{fig:INSdata_LSWT}~(e)), at 6.5\,T, we can clearly resolve two gapped dispersive modes crossing at $l=-0.25$. The minimum/maximum of the dispersions is at $l=-0.5$. At 7.6\,T, it appears that one dispersive mode extends down to the elastic line at $l=-0.5$. Additional weak scattering intensity is seen at $\sim 0.8$\,meV energy transfer. At 8.8\,T, we can observe a single dispersive mode with a minimum at $l=-0.5$ and close to the elastic line. These findings are similar for the cuts along ($0$, $0$, $l$) (Fig.~\ref{fig:INSdata_LSWT}~(f)) but with the details of the spectra less clearly resolved. At 7.6\,T (8.8\,T), we additionally observe a spot of enhanced scattering intensity at $\sim 0.8$\,meV (1\,meV) energy transfer close to $l=0$. Further, at 8.8\,T, we can observe the single dispersive mode extending down to the elastic line at $l=-0.5$. Generally, at 8.8\,T, the minimum at $l = -0.5$ is in accordance with the results along $k$ where for those slices with $l=-0.5$ intensity was mainly observed close to the elastic line.

Before proceeding with the analysis of the $E$-$\mathbf{Q}$ slices, in the given context, we briefly discuss our applied-field INS data along $k$ and $l$ in view of the zero-field data from Ref.~\cite{Rule2017}. Recorded in zero magnetic field and at 500\,mK, those previous INS data captured the spin-wave excitations of the incommensurate, elliptical helix with propagation vector $\mathbf{q} = (0, 0.186, 0.5)$. Along ($0$, $k$, $0.5$), multiple modes collapsing at the elastic line at $k = \pm 0.186$ were observed. In the present experiment, in the field-induced phase IV with a commensurate magnetic propagation vector $\mathbf{q} = (0, 0, 0.5)$ at 6.5\,T, we observe two gapped modes along $k$. This transformation of the spectrum reflects the change to (canted) commensurate antiferromagnetism of linarite. The dispersion along ($0$, $-k$, $-0.25$) features two shallow minima at $k \sim \pm 0.15$. Although recorded in a commensurate magnetic state, this reflects remnant features of the incommensurability in the spin dynamics of linarite---imposed by the competing exchanges $J_1 < 0$ and $J_2 > 0$ along $b$. The shallow minima along ($0$, $-k$, $-0.25$) persist up to the highest experimental field of 8.8\,T. Along ($0$, $0.186$, $l$), in zero magnetic field~\cite{Rule2017}, dispersing modes collapsing at the elastic line at $l = \pm 0.5$ were observed resulting in a rather complicated spin-wave spectrum. By applying a magnetic field, in the present study, the spectrum is transformed successively to a rather simple one with a single mode at 8.8\,T featuring a minimum at $l =-0.5$. This reflects an antiferromagnet being driven into a field polarized state and thus supports the predominance of antiferromagnetic exchange interaction along the $c$ axis.

We now proceed with the analysis of the $E$-$\mathbf{Q}$ slices along $h$ (Fig.~\ref{fig:INSdata_LSWT}~(g)--(h)). As for the $l$ direction, these data allow investigation of the interchain couplings. Studies of the spin excitations along the $h$ direction, however, have not yet been reported for linarite. Along ($h$, $0$, $-0.5$) (Fig.~\ref{fig:INSdata_LSWT}~(g)), at 6.5\,T, we observe dispersion of two modes indicating that weak interchain couplings out of the $bc$ plane are present. The 6.5\,T-data sliced along $l$ favor a picture of the upper mode along ($h$, $0$, $-0.5$) being rather flat while the second disperses. At 7.6\,T, along the same reciprocal direction, we observe less distinct dispersion about $0.8$\,meV. At 8.8\,T, we observe weak dispersion about 1\,meV energy transfer along ($h$, $0$, $-0.5$) for $h < -0.5$ as well as enhanced scattering intensity close to the elastic line for $h > -0.5$. We note that the essence of this spectrum is not fully resolved. Now turning to Fig.~\ref{fig:INSdata_LSWT}~(h), along ($h$, $0$, $-0.25$), a broadened dispersion is evident under a field of 6.5\,T. The results at 7.6 and 8.8\,T resemble each other along ($h$, $0$, $-0.25$) with almost no dispersion present between $h=-1$ and $h=0$ ($E \sim 0.8$\,meV / $1$\,meV). This is a curious observation as we know from the measurements under a field of 6.5\,T that weak interchain couplings out of the $bc$ plane must be present. We will return to these findings for the 7.6 and 8.8\,T-slices in Sec.~\ref{sec:LSWT} in the light of LSWT results.

When globally assessing our INS results we note that in addition to the dominant exchange couplings $J_1$ and $J_2$ we have resolved residual exchange couplings forming a very weak 3D network of chains---likely responsible for the magnetic ordering of linarite at $T_\mathrm{N}$. From the data recorded at 6.5 and 8.8\,T we can mostly extract the spin-wave dispersions, whereas those recorded at 7.6\,T are less intense (\textit{e.g.}, along ($0$, $k$, $-0.5$), ($0.5$, $0$, $l$) and ($0$, $0$, $l$)). This observation is consistent with the unusual character of phase V: a previous NMR study in this field region suggested phase separation into the SDW phase and a phase without detectable dipolar long-range order~\cite{Willenberg2016}. From the field-dependent drop of the integrated intensity of magnetic Bragg peaks measured at the phase boundary IV--V at 1.3\,K~\cite{Heinze2019}, we can roughly estimate that both environments are equally contributing to the phase-separated phase V. Thus, in the INS data, the intensity reduction in phase V is reasonable on the basis of previous experiments. At 8.8\,T, along ($-0.5$, $0$, $l$), we curiously seem to observe a flat as well as a (weak) dispersive mode. This appears unusual for a saturated state. At 7.6 and 8.8\,T, for the slices along ($h$, $0$, $-0.5$), the essence of the spectra is not fully resolved. We will discuss these findings later in Sec.~\ref{sec:LSWT}. By looking closely at the $E$-$\mathbf{Q}$ slice along ($0$, $0$, $l$) at 8.8\,T, there is no clear evidence of a Goldstone mode close to $\mathbf{Q} = 0$; for a 2D nematic state, the Goldstone mode should be isotropically observable~\cite{Starykh2014}. To deepen our understanding of the spin excitations of linarite, we will in a next step model the INS data on the basis of DDMRG and LSWT.

\section{Spin-wave analysis}
\label{sec:LSWT}

For a parametrization of our present set of in-field INS data, we have carried out different modeling approaches. First, we have reanalyzed the spin excitation spectra recorded along $k$ and $l$ using DDMRG~\cite{Jeckelmann2002}, now optimizing the coupling parameter set $J_1$--$J_3$ from Ref.~\cite{Rule2017} on the basis of the applied-field INS data (details are given in Ref.~\cite{SI}). Furthermore, we have applied LSWT on the field-induced phases IV and V as well as the paramagnetic state close to field-polarization. In the past, LSWT modeling of the spin-wave excitation spectra in phase I produced already a very good agreement between experiment and theory~\cite{Rule2017}. Following the procedure of that study, the present calculations were carried out using the Matlab library SpinW~\cite{Toth2015}. As this program package uses (quasi)classical numerical methods, we present in the following a corresponding description of linarite with the awareness that it does not fully account for the quantum nature of this system. Still, applying LSWT to model the in-field INS data has proven to be a reasonable approach for linarite as discussed in the previous study~\cite{Rule2017}. The spin-wave analysis hence enables us to asses which features of the spin excitation spectra can be explained as spin waves, in turn facilitating the search for features of the spectra which cannot.

\begin{figure}[b!]
\includegraphics[width=0.85\linewidth]{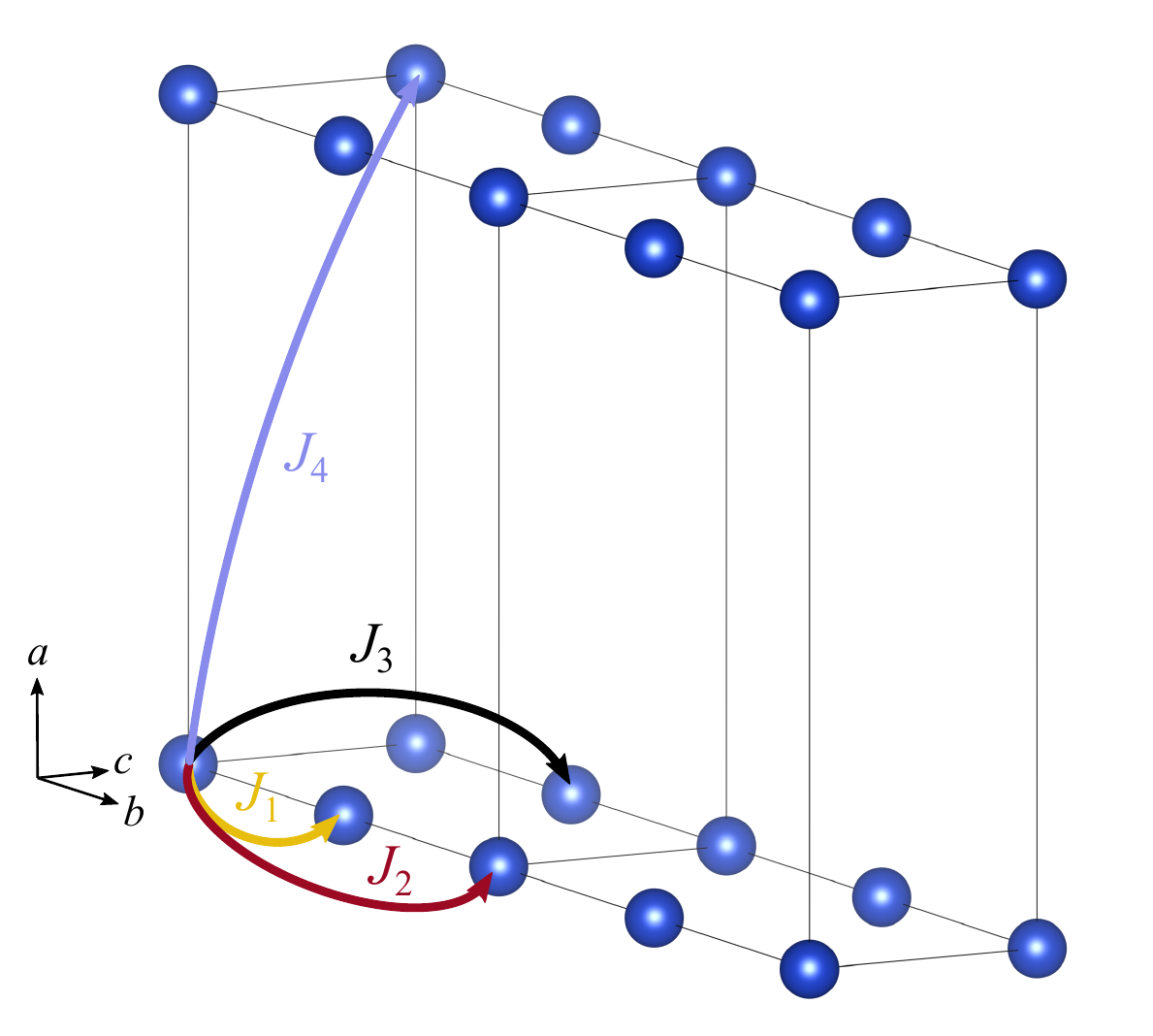}
\caption{Magnetic coupling scheme used for the LSWT calculations consisting of the dominant intrachain couplings $J_1$ and $J_2$ as well as the two residual interchain couplings $J_3$ and $J_4$; the magnetic Cu ions are represented by the blue spheres. Solid lines indicate the structural unit cell.}
\label{fig:couplingscheme}
\end{figure}

Initially, for the calculations in the present study, the exchange interactions obtained from the previous zero-field INS experiment were used: $J_1 = -114$\,K, $J_2 = 37$\,K and $J_3 = 4$\,K~\cite{Rule2017}, as visualized in Fig.~\ref{fig:couplingscheme}. However, as detailed in this Section, for all field-induced phases small modifications to the exchange parameters had to be applied to obtain satisfactory descriptions of the in-field INS data. We refer to the Supplemental Material~\cite{SI} for a comparison between the LSWT calculations using the initial and the optimized set of exchange parameters.

Beyond the results from Ref.~\cite{Rule2017}, in the present study dispersive modes along $h$ were observed in the INS data (Fig.~\ref{fig:INSdata_LSWT} (g)--(h)). Since no estimates for the respective magnetic exchanges in this direction had been made, we performed density-functional-theory (DFT) band-structure calculations using the full-potential code FPLO version 18~\cite{Koepernik1999}. Nonmagnetic calculations were done on a $k$-mesh of $24\times14\times30$ points (2896 points in the irreducible wedge) using the generalized gradient approximation~\cite{Perdew1996} for the exchange and correlation potential. By projecting the half-filled bands onto the Cu $3d_{x^2-y^2}$ states~\cite{SI}, we constructed an effective one-orbital model whose parameters---transfer integrals $t_{ij}$---can be further mapped onto a Hubbard model with an effective onsite repulsion $U_{\text{eff}}$. The corresponding antiferromagnetic exchange $J_{ij}^{\text{AF}}$ was estimated in second-order perturbation theory as $4t_{ij}^2/U_{\text{eff}}$. In this way, we find only one residual coupling out of the $bc$ plane, labeled $J_4$. It corresponds to a diagonal bond running along $a$ and $c$ (Fig.~\ref{fig:couplingscheme}). Taking $U_{\text{eff}}$\,=\,4.5\,eV and adopting $t_4$ from our effective one-orbital model, we arrive at $J_4^{\text{AF}}$\,$\simeq$\,0.13\,K. All other transfer integrals out of the $bc$ plane give rise to minute exchanges that are at least one order of magnitude smaller than $J_4^{\text{AF}}$. In view of previous experimental results, the magnetic propagation vector of linarite, which is of general form $\mathbf{q}=(0, k_y, 0.5)$ in all phases~\cite{Willenberg2012,Willenberg2016}, further supports the DFT finding: A diagonal antiferromagnetic bond in the $ac$ plane stabilizes a ferromagnetic alignment along $a$ with an antiferromagnetic alignment along $c$.

Having established the minimal spin model, we computed its magnon spectra. LSWT calculations were first carried out for a fully field polarized state with a magnetic propagation vector ${\bf q} = (0, 0, 0)$ in a magnetic field of 8.8\,T. These calculations ought to be compared to the INS results under a magnetic field of 8.8\,T, \textit{i.e.}, between the boundary of phase V and the saturation field $\mu_0 H_{\rm sat} = 9.64$\,T~\cite{Heinze2019}. Here, the magnetic character is that of an almost fully field polarized paramagnet~\cite{linarite_Hsat}. Essentially, our experimental observations in 8.8\,T are reproduced on the basis of LSWT with an optimized set of exchange parameters: $J_1 = -114$\,K, $J_2 = 30$\,K and $J_3 = 2.9$\,K and $J_4 = 0.5$\,K~\cite{comment_J1J2}. The parameter set $J_1$--$J_4$ was subsequently applied to phases IV and V. 

Phase IV contains a commensurate magnetic state with a propagation vector ${\bf q} = (0, 0, 0.5)$ where the magnetic moments lie in the $ac$ plane with an angle of $-27^{\circ}$ between the moment direction and the $a$ axis~\cite{Willenberg2016}. For the LSWT calculations, the magnetic structure from the refinement in Ref.~\cite{Willenberg2016} was set as initial magnetic structure. Then, the spins were canted out of the $ac$ plane by $+33^{\circ}$ to account for the effect of the external magnetic field $\mathbf{H} \parallel b$ axis. At this canting angle, the best matching between the INS data and the calculations was obtained. The LSWT calculations were carried out for a supercell matching the magnetic unit cell. The magnetic phase V was modeled as a longitudinal SDW with a magnetic propagation vector ${\bf q} = (0, k_y, 0.5)$ according to the refinement of the magnetic structure carried out in Ref.~\cite{Willenberg2016}. The present INS measurements at 1.3\,K and 7.6\,T were carried out in the high-field region of phase V where $k_y$ is almost field and temperature independent~\cite{Heinze2019}. Following these findings, $k_y = 1/11$ was set for the calculations. Moreover, to account for the effect of the magnetic field on the spin structure, the spin density was shifted towards positive values to obtain a magnetic state with $M = 0.8\,M_\mathrm{sat}$ (see magnetization measurements in Ref.~\cite{Wolter2012}). The calculations were carried out for a supercell matching the magnetic unit cell. 

In the second, fourth and sixth rows of Fig.~\ref{fig:INSdata_LSWT}, we present the LSWT results for the magnetic states described above. They are presented as color-coded spin-wave dispersions plotted as energy transfer over $(h, k, l)$. The calculated spectra were convoluted with a Gaussian function of 0.1\,meV width to simulate a finite energy resolution comparable to the experimental data. We provide the SpinW scripts used in Ref.~\cite{SI}.

The LSWT results describe well the spin-wave excitation spectra along $k$ within our accessible range in reciprocal space and along the energy transfer axis (Fig.~\ref{fig:INSdata_LSWT}~(a)--(c)). At 6.5\,T, our model describes the two distinct modes observed along ($0.5$, $-k$, $-0.5$) and ($0$, $-k$, $-0.5$) as well as the two overlapping modes in the experimental data along ($0$, $-k$, $-0.25$). For phase V, at 7.6\,T, the agreement is more limited. LSWT here predicts a complex spectrum consisting of multiple modes---most of them very weak---and which are not entirely resolved in the present experiment; likely due to the phase separation. At 8.8\,T, the LSWT calculations support the experimental data sliced along ($0.5$, $-k$, $-0.5$) and ($0$, $-k$, $-0.5$), where the single dispersive mode is mainly close to the elastic line within the accessible $k$ range. LSWT further produces a single gapped mode along ($0$, $-k$, $-0.25$), but at a lower energy transfer than observed experimentally. In the INS data, we further observe a strong broadening of this mode (Fig.~\ref{fig:INSdata_LSWT}~(b)), which we found to occur mainly due to the integration with $l \in [-0.35,-0.15]$ for producing the corresponding $E$-$\mathbf{Q}$ maps from the INS data. In this $l$ range, the 8.8\,T-mode is strongly dispersive along $l$ and thus the corresponding slice covers a broader energy transfer range. The same applies for the INS data along ($0$, $-k$, $-0.25$) at 7.6\,T. We present an exemplary demonstration of this broadening in Ref.~\cite{SI} by performing a resolution analysis in Horace/SpinW~\cite{Tobyfit}, which mimics the slicing of experimental INS data to obtain $E$-$\mathbf{Q}$ maps, and also takes into account the finite size, angular divergence and time spread of the neutron beam using a Monte Carlo sampling method.

When comparing experiment and LSWT results for the directions perpendicular to the (reciprocal) chain direction, we find good agreement with our experimental data for the $l$ direction (Fig.~\ref{fig:INSdata_LSWT}~(d)--(f)). Here, we can properly describe the dispersing and crossing of the two gapped modes at 6.5\,T along all three directions as observed in the INS experiment. At 7.6\,T, agreement for the directions ($0.5$, $0$, $l$) and ($0$, $0$, $l$) is found, where the INS data suggest that the most intense dispersive mode extends down to the elastic line at $l=-0.5$. However, the experimental intensity of this mode is very low---in agreement with the phase separation of phase V as discussed in Sec.~\ref{sec:ins_results}. Differently, along ($-0.5$, $0$, $l$) a flat band was observed experimentally, but LSWT here predicts a dispersive mode analogously to the directions ($0.5$, $0$, $l$) and ($0$, $0$, $l$). The case is similar for the field polarized state: At 8.8\,T, the LSWT results along ($0.5$, $0$, $l$) and ($0$, $0$, $l$) also match well the experimental observations with the single dispersive mode extending down to the elastic line at $l = -0.5$. Again, along ($-0.5$, $0$, $l$) the experimental data predominantly suggest a flat band, while LSWT predicts dispersion. We will address this observation later in this Section.

Along $h$ (Fig.~\ref{fig:INSdata_LSWT}~(g)--(h)), we find the best agreement at 6.5\,T, where LSWT can widely describe the two modes along ($h$, $0$, $-0.5$) with an amplitude comparable to experiment for the lower mode, which disperses more strongly. For small $h$, the experimentally observed dispersion appears a little skewed. At 6.5\,T, there is less agreement along ($h$, $0$, $-0.25$), where a similar broadening occurs as for the INS data along ($0$, $-k$, $-0.25$)~\cite{SI}. At 7.6\,T, the agreement between LSWT and experimental results is quite limited: Experimentally, along ($h$, $0$, $-0.5$), we observe weakly dispersive scattering intensity about $\sim 0.8$\,meV whereas LSWT mainly predicts intensity close to the elastic line. Along ($h$, $0$, $-0.25$), although we know from the results at 6.5\,T that weak exchange interactions out of the $bc$ plane must be present, we experimentally observe an (almost) flat mode at $\sim 0.8$\,meV energy transfer at 7.6\,T. The case is similar at 8.8\,T: Along ($h$, $0$, $-0.5$), for $h < -0.5$ there is weakly dispersive scattering intensity about $\sim 1$\,meV, while for $h > -0.5$ we observe scattering intensity close to the elastic line. Here, a weakly dispersive mode close to the elastic line would be in accordance with the LSWT results. Along ($h$, $0$, $-0.25$), as for the results at 7.6\,T, we experimentally observe an almost flat mode at $\sim 1$\,meV at 8.8\,T. This is in disagreement with the LSWT results predicting dispersion at lower energy transfer.

One observation standing out of the spin-wave analysis is that of the apparently flat bands in the experimental spin excitation spectra along ($-0.5$, $0$, $l$) and ($h$, $0$, $-0.25$), and which are not reproduced by LSWT. Also, along ($h$, $0$, $-0.5$), we observe weakly $\mathbf{Q}$-dependent scattering intensity about $\sim 1$\,meV at 8.8\,T for $h < -0.5$, in disagreement with the LSWT results. As the seemingly flat bands cannot be explained as spin waves within our model with exchange parameters $J_1$--$J_4$ but still feature a field-dependent shift towards larger energy transfers, we present an analysis of the intensity distribution at constant energy transfers in Ref.~\cite{SI}. As yet, however, the origin of the observed intensity distribution is unknown but it appears to be intrinsic to linarite.

Summarizing our LSWT-based spin-wave analysis, we can describe the coupling out of the $bc$ plane, which is of the order of $\lesssim 1$\,K, \textit{i.e.}, $\lesssim 1\%$ of $J_1$. Our findings reflect the dramatically different energy scales between the dominant and the residual couplings present in linarite. We have for the first time addressed and resolved the relevance of couplings into all three dimensions for a strongly frustrated quantum spin chain.

\section{Discussion}
\label{sec:discussion}

Altogether, we have carried out an extreme-environment INS experiment on the \textit{ff} $J_1$-$J_2$ chain linarite using magnetic fields up to 8.8\,T and temperatures as low as 500\,mK under the condition of a challenging sample mass of $\sim 150$\,mg. Together with previous INS measurements recorded in zero magnetic field~\cite{Rule2017}, we have for the first time measured the spin excitation spectra of linarite in the entire field range of its magnetic phase diagram, \textit{i.e.}, the magnetic phases I, IV, V and the field polarized state close to saturation. The INS experiment---motivated by the search for spin-multipolar states close to saturation---has led us first to a parametrization of the applied-field INS data by means of DDMRG and LSWT using the parameter set $J_1$--$J_3$~\cite{Rule2017}. On the basis of our experimental data, however, we conclude that an additional exchange path out of the $bc$ plane is needed to explain the dispersive spin-wave modes along $h$. A DFT-based analysis supports this finding and identifies $J_4$, a diagonal bond within the $ac$ plane, to be the leading exchange out of the $bc$ plane albeit with a size of only $J_4^\text{AF} \sim 0.1$\,K. The DFT calculations exclude the relevance of alternative exchange paths out of the $bc$ plane.

We have introduced this very weak exchange $J_4$ in the magnetic model of linarite. Together with the dominant intrachain couplings $J_1 = -114$\,K and $J_2 = 30$\,K ($\alpha = -3.8$) as well as an (effective) residual interchain coupling $J_3 = 2.9$\,K, we have derived an effective 3D coupling scheme. For all magnetic fields from zero field up to saturation, we have parametrized our INS data using LSWT, hereby starting with the field polarized state beyond phase V. Interestingly, the DFT calculations have also identified the interchain exchange perpendicular to the chains ($d_{\text{Cu..Cu}}$\,=\,4.6829\,\r{A}) to be the leading interchain exchange within the $bc$ plane, while the diagonal exchange $J_3$ used in the present work and previous studies was found to be significantly smaller~\cite{SI}. Revisiting our LSWT calculations, we have therefore compared the spin-wave spectra calculated for a field polarized state at 8.8\,T for three different interchain coupling schemes---on our experimental basis, however, we cannot distinguish between the three models~\cite{SI}. While we can neither identify nor exclude a direct bond along $c$ on basis of our INS data, we note that the topology of the interchain couplings might affect a possible presence of spin nematicity in linarite.

In view of the spin-wave analysis, our experimental data pointed to a couple of issues---especially concerning the spin dynamics in phase V and in the magnetic state close to saturation. Here, the LSWT-based analysis yielded some discrepancies to the INS data, which we discuss in the following: (1) The experimental spin-wave spectra recorded in phase V appear less intense compared to LSWT and to the spectra recorded in the other two phases. As discussed in Sec.~\ref{sec:ins_results}, this discrepancy is consistent with phase separation observed previously by means of NMR~\cite{Willenberg2016}. On the basis of previous neutron diffraction results~\cite{Heinze2019}, we can estimate that both phase separated environments roughly contribute equally in this field region yielding an explanation for the reduction of the INS intensity. (2) In phase V and beyond, we observed flat-band features in the experimental spin-wave spectra recorded along ($-0.5$, $0$, $l$) and ($h$, $0$, $-0.25$), and which are not reproduced by means of LSWT. We were able to identify the origin of these seemingly flat bands in our INS data as being an intensity pattern at constant energy transfer~\cite{SI}. As yet, however, the physical origin of these features in $I(\mathbf{Q},E)$ is unknown---although apparently intrinsic to linarite---and requires further investigation. These results point to us that phase V and its peculiarities are not yet fully understood. The same applies for the field region between phase V and the saturation. (3) We found some of the discrepancies of enhanced broadening between INS data and LSWT results being caused by the slicing of the INS data (especially in the spectra with fixed $l=-0.25$). We were able to mimic the slicing for the LSWT calculations using Horace/SpinW and thus uncovered the origin of the broadening being mostly a resolution issue~\cite{SI}.

Considering these points, we have---despite using a sample with a small magnetic moment and very low sample mass---successfully assessed the weak 3D coupling scheme of linarite, which is an important step towards understanding its magnetic behavior. In high magnetic fields the modeling of the spin waves of linarite by means of LSWT works reasonably well as already proven in zero magnetic field~\cite{Rule2017}. We note that in our high-field INS data we do not observe evidence for a Goldstone mode close to $\mathbf{Q} = 0$. If a nematic state exists in linarite at this temperature and field, its signature in the INS data might simply be too weak to be observed in the recorded spin excitation spectra. In our experiment, predominantly, the dipolar character of an (almost) field polarized frustrated spin chain is reflected. In result, proving the existence of a spin-multipolar state in a \textit{ff} $J_1$-$J_2$ chain remains an interesting and at the same time highly challenging task for future investigations.

\begin{acknowledgments}
Experiments at the ISIS Neutron and Muon Source were supported by a beamtime allocation RB1910225 from the Science and Technology Facilities Council. We acknowledge fruitful discussions with U.~K.~R\"{o}{\ss}ler, J.~S.~Gardner and N.~Shannon and thank U. Nitzsche for technical assistance. This work has been supported in part by the Deutsche Forschungs\-gemeinschaft (DFG) under Contract Nos.~WO1532/3-2 and SU229/9-2. A.~U.~B.~W.~and S.~N.~acknowledge financial support from the DFG through the Collaborative Research Center SFB 1143 (project-id 247310070). L.~H.~acknow\-ledges financial support by the Deutscher Akademischer Austauschdienst (DAAD) through a short-term scholarship for PhD students and thanks ANSTO for kind hospitality. The crystal structures presented in Fig.~\ref{fig:phasediagram}~(a)--(b), Fig.~\ref{fig:couplingscheme} and Fig.~S8~\cite{SI} were drawn using the program VESTA 3~\cite{VESTA}.
\end{acknowledgments}


\begin{thebibliography}{99}                                                                                               
\bibitem{Lacroix2011} C. Lacroix, P. Mendels, and F. Mila, \textit{Introduction to Frustrated Magnetism}, Springer Series in Solid-State Sciences (Springer-Verlag, Berlin, 2011).
\bibitem{Broholm2020} C. Broholm, R. J. Cava, S. A. Kivelson, D. G. Nocera, M. R. Norman, T. Senthil, Quantum Spin Liquids, Science \textbf{367}, eaay0668 (2020).
\bibitem{Rule2011} K. C. Rule, D. A. Tennant, J.-S. Caux, M. C. R. Gibson, M. T. F. Telling, S. Gerischer, S. Süllow, and M. Lang, Dynamics of azurite Cu$_3$(CO$_3$)$_2$(OH)$_2$ in a magnetic field as determined by neutron scattering, Phys. Rev. B \textbf{84}, 184419 (2011).
\bibitem{Yoshida2017} M. Yoshida, K. Nawa, H. Ishikawa, M. Takigawa, M. Jeong, S. Kr\"amer, M. Horvati\'c, C. Berthier, K. Matsui, T. Goto, S. Kimura, T. Sasaki, J. Yamaura, H. Yoshida, Y. Okamoto, and Z. Hiroi, Spin dynamics in the high-field phases of volborthite, Phys. Rev. B \textbf{96}, 180413(R) (2017).
\bibitem{Zhang2020} H. Zhang, Z. Zhao, D. Gautreau, M. Raczkowski, A. Saha, V. O. Garlea, H. Cao, T. Hong, H. O. Jeschke, S. D. Mahanti, T. Birol, F. F. Assaad, and X. Ke, Coexistence and Interaction of Spinons and Magnons in an Antiferromagnet with Alternating Antiferromagnetic and Ferromagnetic Quantum Spin Chains, Phys. Rev. Lett. \textbf{125}, 037204 (2020).

\bibitem{Chubukov1991} A. V. Chubukov, Chiral, nematic, and dimer states in quantum spin chains, Phys. Rev. B \textbf{44}, 4693(R) (1991).
\bibitem{Hikihara2008} T. Hikihara, L. Kecke, T. Momoi, and A. Furusaki, Vector chiral and multipolar orders in the spin-$1/2$ frustrated ferromagnetic chain in magnetic field, Phys. Rev. B \textbf{78}, 144404 (2008).
\bibitem{Sudan2009} J. Sudan, A. L\"uscher, and A. M. L\"auchli, Emergent multipolar spin correlations in a fluctuating spiral: The frustrated ferromagnetic spin-$1/2$ Heisenberg chain in a magnetic field, Phys. Rev. B \textbf{80}, 140402(R) (2009).
\bibitem{HeidrichMeisner2009} F. Heidrich-Meisner, I. P. McCulloch, and A. K. Kolezhuk, Phase diagram of an anisotropic frustrated ferromagnetic spin-$1/2$ chain in a magnetic field: A density matrix renormalization group study, Phys. Rev. B \textbf{80}, 144417 (2009).
\bibitem{Zhitomirsky2010}  M. E. Zhitomirsky and H. Tsunetsugu, Magnon pairing in quantum spin nematic, Europhys. Lett. \textbf{92}, 37001 (2010).
\bibitem{Onishi2015} H. Onishi, Magnetic Excitations of Spin Nematic State in Frustrated Ferromagnetic Chain, J. Phys. Soc. Jpn. \textbf{84}, 083702 (2015).
\bibitem{Balents2016} L. Balents and O. A. Starykh, Quantum Lifshitz Field Theory of a Frustrated Ferromagnet, Phys. Rev. Lett. \textbf{116}, 177201 (2016).
\bibitem{Sato2013} M. Sato, T. Hikihara, and T. Momoi, Spin-Nematic and Spin-Density-Wave Orders in Spatially Anisotropic Frustrated Magnets in a Magnetic Field, Phys. Rev. Lett. \textbf{110}, 077206 (2013).
\bibitem{Jinterchain_and_DM} By including interchain coupling in Eq.~(\ref{eq:Hamiltonian}), 3D ordered incommensurate SDW or nematic states can be stabilized whereas the size of the nematic phase depends on the size of the interchain interactions~\cite{Sato2013}. A staggered DM interaction, as allowed in linarite merely on the $J_1$ bond, can lead to a characteristic behavior with an asymptotical saturation of the magnetization. As this is not observed for linarite (see for instance Fig.~3 in Ref.~\cite{Wolter2012}), DM interaction is considered a secondary effect at present stage.

\bibitem{Sato2009} M. Sato, T. Momoi, and A. Furusaki, NMR relaxation rate and dynamical structure factors in nematic and multipolar liquids of frustrated spin chains under magnetic fields, Phys. Rev. B \textbf{79}, 060406(R) (2009).
\bibitem{Furuya2017} S. C. Furuya, Angular dependence of electron spin resonance for detecting the quadrupolar liquid state of frustrated spin chains, Phys. Rev. B \textbf{95}, 014416 (2017).
\bibitem{Flavia2018} F. B. Ramos, S. Eli\"ens, and R. G. Pereira, Dynamical structure factors in the nematic phase of frustrated ferromagnetic spin chains, Phys. Rev. B \textbf{98}, 094431 (2018). 

\bibitem{Enderle2005} M. Enderle, C. Mukherjee, B. F{\aa}k, R. K. Kremer, J.-M. Broto, H. Rosner, S.-L. Drechsler, J. Richter, J. Malek, and A. Prokofiev, Quantum helimagnetism of the frustrated spin-chain LiCuVO$_4$, Europhys. Lett. \textbf{70}, 237 (2005).
\bibitem{Enderle2010} M. Enderle, B. F{\aa}k, H.-J. Mikeska, R. K. Kremer, A. Prokofiev, and W. Assmus, Two-Spinon and Four-Spinon Continuum in a Frustrated Ferromagnetic Spin-$1/2$ Chain, Phys. Rev. Lett. \textbf{104}, 237207 (2010).
\bibitem{Nishimoto2011} S. Nishimoto, S.-L. Drechsler, R. Kuzian, J. Richter, J. Málek, M. Schmitt, J. van den Brink, and H. Rosner, The strength of frustration and quantum fluctuations in LiVCuO$_4$, Europhys. Lett. \textbf{98}, 37007 (2012).

\bibitem{Wolter2012} A. U. B. Wolter, F. Lipps, M. Sch\"apers, S.-L. Drechsler, S. Nishimoto, R. Vogel, V. Kataev, B. B\"uchner, H. Rosner, M. Schmitt, M. Uhlarz, Y. Skourski, J. Wosnitza, S. S\"ullow, and K. C. Rule, Magnetic properties and exchange integrals of the frustrated chain cuprate linarite PbCuSO$_4$(OH)$_2$, Phys. Rev. B \textbf{85}, 014407 (2012).
\bibitem{Rule2017} K. C. Rule, B. Willenberg, M. Sch\"apers, A. U. B. Wolter, B. B\"uchner, S.-L. Drechsler, G. Ehlers, D. A. Tennant, R. A. Mole, J. S. Gardner, S. S\"ullow, and S. Nishimoto, Dynamics of linarite: Observations of magnetic excitations, Phys. Rev. B \textbf{95}, 024430 (2017).
\bibitem{Cemal2018} E. Cemal, M. Enderle, R. K. Kremer, B. F{\aa}k, E. Ressouche, J. P. Goff, M. V. Gvozdikova, M. E. Zhitomirsky, and T. Ziman, Field-induced States and Excitations in the Quasicritical Spin-$1/2$ Chain Linarite, Phys. Rev. Lett. \textbf{120}, 067203 (2018).

\bibitem{Saul2014} A. Sa\'ul and G. Radtke, Density functional approach for the magnetism of $\beta$-TeVO$_4$, Phys. Rev. B \textbf{89}, 104414 (2014).
\bibitem{Weickert2016} F. Weickert, N. Harrison, B. L. Scott, M. Jaime, A. Leitm\"ae, I. Heinmaa, R. Stern, O. Janson, H. Berger, H. Rosner, and A. A. Tsirlin, Magnetic anisotropy in the frustrated spin-chain compound $\beta$-TeVO$_4$, Phys. Rev. B \textbf{94}, 064403 (2016).

\bibitem{Buettgen2014} N. B\"uttgen, K. Nawa, T. Fujita, M. Hagiwara, P. Kuhns, A. Prokofiev, A. P. Reyes, L. E. Svistov, K. Yoshimura, and M. Takigawa, Search for a spin-nematic phase in the quasi-one-dimensional frustrated magnet LiCuVO$_4$, Phys. Rev. B \textbf{90}, 134401 (2014).
\bibitem{Orlova2017} A. Orlova, E. L. Green, J. M. Law, D. I. Gorbunov, G. Chanda, S. Kr\"amer, M. Horvati\'c, R. K. Kremer, J. Wosnitza, and G. L. J. A. Rikken, Nuclear Magnetic Resonance Signature of the Spin-Nematic Phase in LiCuVO$_4$ at High Magnetic Fields, Phys. Rev. Lett. \textbf{118}, 247201 (2017).
\bibitem{Pregelj2020} M. Pregelj, A. Zorko, D. Ar\v{c}on, M. Klanj\v{s}ek, O. Zaharko, S. Kr\"amer, M. Horvati\'c, and A. Prokofiev, Thermal effects versus spin nematicity in a frustrated spin-$1/2$ chain, Phys. Rev. B \textbf{102}, 081104(R) (2020).

\bibitem{Heinze2019} L. Heinze, G. Bastien, B. Ryll, J.-U. Hoffmann, M. Reehuis, B. Ouladdiaf, F. Bert, E. Kermarrec, P. Mendels, S. Nishimoto, S.-L. Drechsler, U. K. R\"{o}{\ss}ler, H. Rosner, B. B\"uchner, A. J. Studer, K. C. Rule, S. S\"ullow, and A. U. B. Wolter, Magnetic phase diagram of the frustrated spin chain compound linarite PbCuSO$_4$(OH)$_2$ as seen by neutron diffraction and $^1$H-NMR, Phys. Rev. B \textbf{99}, 094436 (2019).
\bibitem{Willenberg2016} B. Willenberg, M. Sch\"{a}pers, A. U. B. Wolter, S.-L. Drechsler, M. Reehuis, J.-U. Hoffmann, B. B\"{u}chner, A. J. Studer, K. C. Rule, B. Ouladdiaf, S. S\"{u}llow, and S. Nishimoto, Complex Field-Induced States in Linarite PbCuSO$_4$(OH)$_2$ with a Variety of High-Order Exotic Spin-Density Wave States, Phys. Rev. Lett. \textbf{116}, 047202 (2016).
\bibitem{Gotovko2019} S. K. Gotovko, L. E. Svistov, A. M. Kuzmenko, A. Pimenov, and M. E. Zhitomirsky, Electron spin resonance in spiral antiferromagnet linarite: Theory and experiment, Phys. Rev. B \textbf{100}, 174412 (2019).
\bibitem{Willenberg2012} B. Willenberg, M. Sch\"{a}pers, K. C. Rule, S. S\"{u}llow, M. Reehuis, H. Ryll, B. Klemke, K. Kiefer, W. Schottenhamel, B. B\"{u}chner, B. Ouladdiaf, M. Uhlarz, R. Beyer, J. Wosnitza, and A. U. B. Wolter, Magnetic Frustration in a Quantum Spin Chain: The Case of Linarite PbCuSO$_4$(OH)$_2$, Phys. Rev. Lett. \textbf{108}, 117202 (2012).
\bibitem{Schaepers2013} M. Sch\"{a}pers, A. U. B. Wolter, S.-L. Drechsler, S. Nishimoto, K.-H. M\"{u}ller, M. Abdel-Hafiez, W. Schottenhamel, B. B\"{u}chner, J. Richter, B. Ouladdiaf, M. Uhlarz, R. Beyer, Y. Skourski, J. Wosnitza, K. C. Rule, H. Ryll, B. Klemke, K. Kiefer, M. Reehuis, B. Willenberg, and S. S\"{u}llow, Thermodynamic properties of the anisotropic frustrated spin-chain compound linarite PbCuSO$_4$(OH)$_2$, Phys. Rev. B \textbf{88}, 184410 (2013).
\bibitem{Schaepers2014} M. Sch\"{a}pers, H. Rosner, S.-L. Drechsler, S. S\"{u}llow, R. Vogel, B. B\"{u}chner, and A. U. B. Wolter, Magnetic and electronic structure of the frustrated spin-chain compound linarite PbCuSO$_4$(OH)$_2$, Phys. Rev. B \textbf{90}, 224417 (2014).
\bibitem{Povarov2016} K. Yu. Povarov, Y. Feng, and A. Zheludev, Multiferroic phases of the frustrated quantum spin-chain compound linarite, Phys. Rev. B \textbf{94}, 214409 (2016).
\bibitem{Feng2018} Y. Feng, K. Yu. Povarov, and A. Zheludev, Magnetic phase diagram of the strongly frustrated quantum spin chain system PbCuSO$_4$(OH)$_2$ in tilted magnetic fields, Phys. Rev. B \textbf{98}, 054419 (2018).
\bibitem{Gillig2021} M. Gillig, X. Hong, P. Sakrikar, G. Bastien, A. U. B. Wolter, L. Heinze, S. Nishimoto, B. B\"uchner, and C. Hess, Thermal transport of the frustrated spin-chain mineral linarite: Magnetic heat transport and strong spin-phonon scattering, Phys. Rev. B \textbf{104}, 235129 (2021).

\bibitem{Smerald2015} A. Smerald, H. T. Ueda, and N. Shannon, Theory of inelastic neutron scattering in a field-induced spin-nematic state, Phys. Rev. B \textbf{91}, 174402 (2015).
\bibitem{Starykh2014} O. A. Starykh and L. Balents, Excitations and quasi-one-dimensionality in field-induced nematic and spin density wave states, Phys. Rev. B \textbf{89}, 104407 (2014).

\bibitem{Effenberger1987} H. Effenberger, Crystal structure and chemical formula of schmiederite, Pb$_2$Cu$_2$(OH)$_4$(SeO$_3$)(SeO$_4$), with a comparison to linarite, PbCu(OH)$_2$(SO$_4$), Miner. Petrol. \textbf{36}, 3 (1987).
\bibitem{Schofield2009} P. F. Schofield, C. C. Wilson, K. S. Knight, and C. A. Kirk, Proton location and hydrogen bonding in the hydrous lead copper sulfates linarite, PbCu(SO$_4$)(OH)$_2$, and caledonite, Pb$_5$Cu$_2$(SO$_4$)$_3$CO$_3$(OH)$_6$, Can. Mineral. \textbf{47}, 649 (2009).
\bibitem{LET} R. I. Bewley, J. W. Taylor, and S. M. Bennington, LET, a cold neutron multi-disk chopper spectrometer at ISIS, Nucl. Instrum. Methods Phys. Res. Sect. A \textbf{637}, 128 (2011).
\bibitem{Rule2018glue} K. C. Rule, R. A. Mole, and D. Yu, Which glue to choose? A neutron scattering study of various adhesive materials and their effect on background scattering, J. Appl. Cryst. \textbf{51}, 6 (2018).
\bibitem{RB1910225} K. C. Rule, L. Heinze, M. D. Le, and S. S\"ullow, STFC ISIS Neutron and Muon Source, https://doi.org/10.5286/ISIS.E.RB1910225 (2019).
\bibitem{SI} See Supplemental Material at (URL will be inserted by publisher) for the presentation of the INS data recorded at 500\,mK as well as constant-energy slices through INS data. Further, additional information on the various accompanying calculations for this work is given: This includes the applied-field DDMRG, the DFT band-structure and the LSWT calculations. The Supplemental Material includes Ref.~\cite{Nishimoto2007}.
\bibitem{Nishimoto2007} S. Nishimoto and M. Arikawa, Dynamics in One-Dimensional Spin Systems -- Density-Matrix Renormalization Group Study, Int. J. Mod. Phys. B \textbf{21}, 2262 (2007).
\bibitem{Mantid} O. Arnold, J. C. Bilheux, J. M. Borreguero, A. Buts, S.I. Campbell, L. Chapon, M. Doucet, N. Draper, R. Ferraz Leal, M.A. Gigg, V.E. Lynch, A. Markvardsen, D. J. Mikkelson, R. L. Mikkelson, R. Miller, K. Palmen, P. Parker, G. Passos, T. G. Perring, P. F. Peterson, S. Ren, M. A. Reuter, A. T. Savici, J. W. Taylor, R. J. Taylor, R. Tolchenov, W. Zhou, J. Zikovsky, Mantid---Data analysis and visualization package for neutron scattering and $\mu$SR experiments, Nucl. Instrum. Methods A \textbf{764}, 156 (2014). 
\bibitem{Horace} R. A. Ewings, A. Buts, M. D. Le, J. van Duijn, I. Bustinduy, T. G. Perring, Horace: Software for the analysis of data from single crystal spectroscopy experiments at time-of-flight neutron instruments, Nucl. Instrum. Methods Phys. Res. Sect. A \textbf{834}, 132 (2016). 
\bibitem{cut_0k-0p25} During the spin-wave analysis and by carrying out a resolution analysis in Horace/SpinW using Tobyfit, we found that the broadening along ($0$, $-k$, $-0.25$) is mainly due to the integration along $l$ when producing the corresponding $E$-$\mathbf{Q}$ slice. We present an exemplary demonstration of the broadening along ($0$, $-k$, $-0.25$) in the Supplemental Material~\cite{SI}. 
\bibitem{Jeckelmann2002} E. Jeckelmann, Dynamical density-matrix renormalization-group method, Phys. Rev. B \textbf{66}, 045114 (2002).
\bibitem{Toth2015} S. Toth and B. Lake, Linear spin wave theory for single-$Q$ incommensurate magnetic structures, J. Phys.: Condens. Matter \textbf{27}, 166002 (2015).
\bibitem{Koepernik1999} K. Koepernik and H. Eschrig, Full-potential nonorthogonal local-orbital minimum-basis band-structure scheme, Phys. Rev. B \textbf{59}, 1743 (1999).
\bibitem{Perdew1996} J. P. Perdew, K. Burke, and M. Ernzerhof, Generalized Gradient Approximation Made Simple, Phys. Rev. Lett. \textbf{77}, 3865 (1996).
\bibitem{linarite_Hsat} From the magnetization measured as function of the magnetic field at $T = 1.8$\,K reported in Ref.~\cite{Wolter2012}, we extract $M/M_\mathrm{sat} \sim 0.95$.
\bibitem{comment_J1J2} We note that with the limited access to the spin-wave dispersion along $k$ and the maximal energy transfer of $\sim 2.5$\,meV, we have a more limited access to the intrachain couplings $J_1$ and $J_2$ than to the interchain couplings $J_3$ and $J_4$. For this reason, we only applied small changes to the initial parameters from Ref.~\cite{Rule2017}, which were determined for an extended $k$ range~\cite{SI}.
\bibitem{Tobyfit} \url{https://pace-neutrons.github.io/Horace/v3.6.2/user_guide/Resolution_convolution.html}
\bibitem{VESTA} K. Momma and F. Izumi, VESTA 3 for three-dimensional visualization of crystal, volumetric and morphology data, J. Appl. Crystallogr. \textbf{44}, 1272 (2011).
\end{thebibliography}
\end{document}